\documentstyle[twoside,psfig]{article}

\newcommand{\bfy}{{\bf y}}
\newcommand{\bfx}{{\bf x}}
\newcommand{\Schrodinger}{Schr\"{o}dinger\ }

\catcode`\@=11
\long\def\@makefntext#1{
\protect\noindent \hbox to 3.2pt {\hskip-.9pt  
$^{{\eightrm\@thefnmark}}$\hfil}#1\hfill}		

\def\thefootnote{\fnsymbol{footnote}}
\def\@makefnmark{\hbox to 0pt{$^{\@thefnmark}$\hss}}	
	
\def\ps@myheadings{\let\@mkboth\@gobbletwo
\def\@oddhead{\hbox{}
\rightmark\hfil\eightrm\thepage}   
\def\@oddfoot{}\def\@evenhead{\eightrm\thepage\hfil
\leftmark\hbox{}}\def\@evenfoot{}
\def\sectionmark##1{}\def\subsectionmark##1{}}

\oddsidemargin=\evensidemargin
\renewcommand{\thefootnote}{\fnsymbol{footnote}}

\newcounter{sectionc}\newcounter{subsectionc}\newcounter{subsubsectionc}
\renewcommand{\section}[1] {\vspace{12pt}\addtocounter{sectionc}{1} 
\setcounter{subsectionc}{0}\setcounter{subsubsectionc}{0}\noindent 
	{\tenbf\thesectionc. #1}\par\vspace{5pt}}
\renewcommand{\subsection}[1] {\vspace{12pt}\addtocounter{subsectionc}{1} 
	\setcounter{subsubsectionc}{0}\noindent 
	{\bf\thesectionc.\thesubsectionc. {\kern1pt \bfit #1}}\par\vspace{5pt}}
\renewcommand{\subsubsection}[1] {\vspace{12pt}\addtocounter{subsubsectionc}{1}
	\noindent{\tenrm\thesectionc.\thesubsectionc.\thesubsubsectionc.
	{\kern1pt \tenit #1}}\par\vspace{5pt}}
\newcommand{\nonumsection}[1] {\vspace{12pt}\noindent{\tenbf #1}
	\par\vspace{5pt}}

\newcounter{appendixc}
\newcounter{subappendixc}[appendixc]
\newcounter{subsubappendixc}[subappendixc]
\renewcommand{\thesubappendixc}{\Alph{appendixc}.\arabic{subappendixc}}
\renewcommand{\thesubsubappendixc}
	{\Alph{appendixc}.\arabic{subappendixc}.\arabic{subsubappendixc}}

\renewcommand{\appendix}[1] {\vspace{12pt}
        \refstepcounter{appendixc}
        \setcounter{figure}{0}
        \setcounter{table}{0}
        \setcounter{lemma}{0}
        \setcounter{theorem}{0}
        \setcounter{corollary}{0}
        \setcounter{definition}{0}
        \setcounter{equation}{0}
        \renewcommand{\thefigure}{\Alph{appendixc}.\arabic{figure}}
        \renewcommand{\thetable}{\Alph{appendixc}.\arabic{table}}
        \renewcommand{\theappendixc}{\Alph{appendixc}}
        \renewcommand{\thelemma}{\Alph{appendixc}.\arabic{lemma}}
        \renewcommand{\thetheorem}{\Alph{appendixc}.\arabic{theorem}}
        \renewcommand{\thedefinition}{\Alph{appendixc}.\arabic{definition}}
        \renewcommand{\thecorollary}{\Alph{appendixc}.\arabic{corollary}}
        \renewcommand{\theequation}{\Alph{appendixc}.\arabic{equation}}
        \noindent{\tenbf Appendix \theappendixc #1}\par\vspace{5pt}}
\newcommand{\subappendix}[1] {\vspace{12pt}
        \refstepcounter{subappendixc}
        \noindent{\bf Appendix \thesubappendixc. {\kern1pt \bfit #1}}
	\par\vspace{5pt}}
\newcommand{\subsubappendix}[1] {\vspace{12pt}
        \refstepcounter{subsubappendixc}
        \noindent{\rm Appendix \thesubsubappendixc. {\kern1pt \tenit #1}}
	\par\vspace{5pt}}

\newcommand{\textlineskip}{\baselineskip=13pt}
\newcommand{\smalllineskip}{\baselineskip=10pt}

\def\eightcirc{
\begin{picture}(0,0)
\put(4.4,1.8){\circle{6.5}}
\end{picture}}
\def\eightcopyright{\eightcirc\kern2.7pt\hbox{\eightrm c}} 

\newcommand{\copyrightheading}[1]
	{\vspace*{-2.5cm}\smalllineskip{\flushright
	{\footnotesize PUPT-1677, BU-CCS-970102 #1}\\
	{\footnotesize quant-ph/9701016}\\
	 }}

\def\abstracts#1#2#3{{
	\centering{\begin{minipage}{4.5in}\baselineskip=10pt\footnotesize
	\parindent=0pt #1\par 
	\parindent=15pt #2\par
	\parindent=15pt #3
	\end{minipage}}\par}} 

\newcommand{\bibit}{\nineit}

\renewenvironment{thebibliography}[1]
        {\frenchspacing
	 \ninerm\baselineskip=11pt
         \begin{list}{\arabic{enumi}.}
        {\usecounter{enumi}\setlength{\parsep}{0pt}     
	 \setlength{\leftmargin 12.7pt}{\rightmargin 0pt} 
         \setlength{\itemsep}{0pt} \settowidth
	{\labelwidth}{#1.}\sloppy}}{\end{list}}

\newcounter{itemlistc}
\newcounter{romanlistc}
\newcounter{alphlistc}
\newcounter{arabiclistc}

\newcommand{\fcaption}[1]{
        \refstepcounter{figure}
        \setbox\@tempboxa = \hbox{\footnotesize Fig.~\thefigure. #1}
        \ifdim \wd\@tempboxa > 5in
           {\begin{center}
        \parbox{5in}{\footnotesize\smalllineskip Fig.~\thefigure. #1}
            \end{center}}
        \else
             {\begin{center}
             {\footnotesize Fig.~\thefigure. #1}
              \end{center}}
        \fi}

\newcommand{\tcaption}[1]{
        \refstepcounter{table}
        \setbox\@tempboxa = \hbox{\footnotesize Table~\thetable. #1}
        \ifdim \wd\@tempboxa > 5in
           {\begin{center}
        \parbox{5in}{\footnotesize\smalllineskip Table~\thetable. #1}
            \end{center}}
        \else
             {\begin{center}
             {\footnotesize Table~\thetable. #1}
              \end{center}}
        \fi}

\def\@citex[#1]#2{\if@filesw\immediate\write\@auxout
	{\string\citation{#2}}\fi
\def\@citea{}\@cite{\@for\@citeb:=#2\do
	{\@citea\def\@citea{,}\@ifundefined
	{b@\@citeb}{{\bf ?}\@warning
	{Citation `\@citeb' on page \thepage \space undefined}}
	{\csname b@\@citeb\endcsname}}}{#1}}

\newif\if@cghi
\def\cite{\@cghitrue\@ifnextchar [{\@tempswatrue
	\@citex}{\@tempswafalse\@citex[]}}
\def\citelow{\@cghifalse\@ifnextchar [{\@tempswatrue
	\@citex}{\@tempswafalse\@citex[]}}
\def\@cite#1#2{{$\null^{#1}$\if@tempswa\typeout
	{IJCGA warning: optional citation argument 
	ignored: `#2'} \fi}}

\def\pmb#1{\setbox0=\hbox{#1}
	\kern-.025em\copy0\kern-\wd0
	\kern.05em\copy0\kern-\wd0
	\kern-.025em\raise.0433em\box0}

\def\fnt#1#2{\footnotetext{\kern-.3em
	{$^{\mbox{\scriptsize #1}}$}{#2}}}

\def\fpage#1{\begingroup
\voffset=.3in
\thispagestyle{empty}\begin{table}[b]\centerline{\footnotesize #1}
	\end{table}\endgroup}

\def\runninghead#1#2{\pagestyle{myheadings}
\markboth{{\protect\footnotesize\it{\quad #1}}\hfill}
{\hfill{\protect\footnotesize\it{#2\quad}}}}
\font\tenrm=cmr10
\font\tenit=cmti10 
\font\tenbf=cmbx10
\font\bfit=cmbxti10 at 10pt
\font\ninerm=cmr9
\font\nineit=cmti9

\font\eightrm=cmr8

\def\qed{\hbox{${\vcenter{\vbox{			
   \hrule height 0.4pt\hbox{\vrule width 0.4pt height 6pt
   \kern5pt\vrule width 0.4pt}\hrule height 0.4pt}}}$}}

\renewcommand{\thefootnote}{\fnsymbol{footnote}}	

\def\bsc{{\sc a\kern-6.4pt\sc a\kern-6.4pt\sc a}}  	
\def\bflatex{\bf L\kern-.30em\raise.3ex\hbox{\bsc}\kern-.14em 
T\kern-.1667em\lower.7ex\hbox{E}\kern-.125em X} 

\begin{document}

\runninghead{Quantum lattice-gas models for the many-body
Schr\"odinger equation} {Quantum lattice-gas models for the many-body
Schr\"odinger equation}

\normalsize\textlineskip
\thispagestyle{empty}
\setcounter{page}{1}

\copyrightheading{}			

\vspace*{0.88truein}

\fpage{1}
\centerline{\bf Quantum lattice-gas models}
\vspace*{0.035truein}
\centerline{\bf for the many-body Schr\"odinger equation\footnote{Talk
given by WT at the Sixth International Conference on Discrete Fluid
Mechanics, Boston University, Boston MA 1996}}
\vspace*{0.37truein}
\centerline{\footnotesize BRUCE M. BOGHOSIAN}
\vspace*{0.015truein}
\centerline{\footnotesize\it Center for Computational Science}
\baselineskip=10pt
\centerline{\footnotesize\it Boston University}
\baselineskip=10pt
\centerline{\footnotesize\it 3 Cummington Street}
\baselineskip=10pt
\centerline{\footnotesize\it Boston, Massachusetts 02215, USA}
\baselineskip=10pt
\centerline{\footnotesize\tt bruceb@bu.edu}
\vspace*{10pt}
\centerline{\normalsize and}
\vspace*{10pt}
\centerline{\footnotesize WASHINGTON TAYLOR IV}
\vspace*{0.015truein}
\centerline{\footnotesize\it Department of Physics}
\baselineskip=10pt
\centerline{\footnotesize\it Joseph Henry Laboratories}
\baselineskip=10pt
\centerline{\footnotesize\it Princeton University}
\baselineskip=10pt
\centerline{\footnotesize\it Princeton, New Jersey 08544, USA}
\baselineskip=10pt
\centerline{\footnotesize\tt wati@princeton.edu}
\vspace*{0.225truein}

\vspace*{0.21truein}
\abstracts{
A general class of discrete unitary models are described whose
behavior in the continuum limit corresponds to a many-body
\Schrodinger equation.  On a quantum computer, these models could be
used to simulate quantum many-body systems with an exponential speedup
over analogous simulations on classical computers.  On a classical
computer, these models give an explicitly unitary and local
prescription for discretizing the \Schrodinger equation.  It is shown
that models of this type can be constructed for an arbitrary number of
particles moving in an arbitrary number of dimensions with an
arbitrary interparticle interaction.  }{}{}



\textheight=7.8truein
\setcounter{footnote}{0}
\renewcommand{\thefootnote}{\alph{footnote}}

\vspace*{1pt}\textlineskip	
\section{Introduction}	
\noindent
In this paper we describe a class of algorithms for
simulating quantum mechanical systems.  These algorithms are very
similar to the lattice-gas automata and lattice Boltzmann models for
hydrodynamics which were discussed in many of the other talks at
this conference.  In the models we will be discussing, however, the
microscopic dynamics is defined by a time-development rule which is
unitary, rather than probability-conserving as in traditional
lattice-gas automata or lattice Boltzmann models.

There are several reasons for believing that these discrete models for
quantum mechanics are interesting.  First, they give an explicitly
unitarity way of discretizing the \Schrodinger equation, so they might
be better behaved numerically than standard finite-difference methods.
Second, they are particularly well suited to implementation on a
quantum computer.  In fact, if someone could build a general purpose
quantum computer\cite{qc}, it would be possible to use these
algorithms to simulate systems of many interacting quantum particles
in exponentially less time than it would take on a classical computer.

David Meyer, who first suggested the term ``quantum lattice-gas
automata\cite{Meyer}'', discussed some aspects of these systems in his
talk, but for completeness we begin by reviewing briefly what a
quantum lattice-gas automaton is, and how such a system is different
from a classical lattice-gas automaton.  We then consider a fairly
generic 1-dimensional quantum lattice system which obeys the
\Schrodinger equation in the continuum limit.  By extending the system
in various ways, we show how multiple particles, higher
dimensionality and interparticle interactions can naturally be
incorporated into the system, so that we end up with a simple
microscopic lattice model which describes an arbitrary system of
interacting nonrelativistic quantum particles.  Finally, we discuss
how these models can be used on a quantum computer to achieve
exponential performance enhancement over an analogous system on a
classical computer.

\section{Quantum lattice-gas automata}
\noindent
The idea of a quantum lattice-gas automaton (QLGA) was first suggested
by David Meyer\cite{Meyer}, who considered such systems in the context
of a multiple-particle Dirac equation in one dimension.  Essentially,
a quantum lattice-gas automaton is very similar to a classical
lattice-gas automaton\cite{lga}.  At each vertex of some lattice
$\Lambda$ there are $m$ quantum bits ({\it q-bits}), which represent
particle occupation numbers.  In a classical lattice-gas automaton
(LGA), there would be $m$ classical bits with values 0 or 1 indicating
the absence or presence of particles with velocities $v_i, 1 \leq i
\leq m$ (see Fig.~\ref{f:QLGA}).
\begin{figure}[htbp]
\vspace*{13pt}
\centerline{\vbox{\hrule width 5cm height0.001pt}}
\begin{center}
\begin{picture}(200,130)(- 100,- 65)
\multiput(-40,40)(80,0){2}{\circle*{5}}
\multiput(-40, -40)(80,0){2}{\circle*{5}}
\multiput(-40,40)(80,0){2}{\vector(1,0){20}}
\multiput(-40,40)(80,0){2}{\vector(0,1){20}}
\multiput(-40,40)(80,0){1}{\vector(0,-1){20}}
\multiput(40, -40)(80,0){1}{\vector(-1,0){20}}
\multiput(-40, -40)(80,0){1}{\vector(0,1){20}}
\multiput(-40, -40)(80,0){2}{\vector(0,-1){20}}
\end{picture}
\end{center}
\centerline{\vbox{\hrule width 5cm height0.001pt}}
\vspace*{13pt}
\fcaption{A state in a 2-dimensional classical LGA with $m = 4$
particle sites at each lattice site.  Arrows indicate occupied sites.
In a quantum lattice-gas automaton, this state would be a basis
vector in a Hilbert space of dimension $2^{4l}$.}
\label{f:QLGA}
\end{figure}
The state space for a classical LGA contains $2^{lm}$ discrete states,
where $l = | \Lambda |$ is the number of lattice sites.  In a QLGA, on
the other hand, the space of allowed states of the system at a fixed
point in time corresponds to a Hilbert space of $lm$ independent
two-state quantum components.  The state space for a single q-bit is a
continuous space parameterized by two complex numbers $\psi_{+},
\psi_{-}$ satisfying $| \psi_{+} |^2 + | \psi_{-} |^2 = 1$.  These
numbers correspond to amplitudes for the presence or absence of a
particle in the QLGA, respectively.  For a system of $n$ q-bits, the
Hilbert space is parameterized by $2^n$ complex numbers
$\psi_{\sigma_1 \cdots \sigma_n}$ where $\sigma_i \in \{+,-\}$.  All states
are normalized so that $\sum_\sigma | \psi_\sigma |^2 = 1$.  In a
quantum lattice-gas system, a natural basis for the Hilbert space is
given by the $2^{lm}$ classical configurations corresponding to
definite particle occupation numbers; the parameters $\psi_\sigma$
correspond to the amplitudes for each state in this basis.

Thus, we see that while the state space for a classical LGA is defined
by an element of a set of $2^{lm}$ states, the state space for a QLGA
is a complex linear vector space of dimension $2^{lm}$, where the
state is restricted to have unit norm.  We now consider the time
development rule for a QLGA.  In a (deterministic) classical LGA, the
time development rule is defined in two steps.  First each particle
advects forward in the direction of its associated velocity vector.
Then, the bits associated with particles at each lattice site are
transformed by acting on the set of $2^m$ possible local states by a
permutation matrix.  Generally, this ``collision  rule'' is defined in
such a way as to conserve particle number and/or momentum, to achieve
the desired hydrodynamic equations in the continuum limit.  In a QLGA,
we break the time development rule into two parts in the same way.
First, we advect the particles by exchanging pairs of q-bits between
adjacent lattice sites in the manner indicated by the velocity
vectors associated with each q-bit.  Next, we act on the set of
q-bits at each lattice site with a fixed collision operator.  Because
the Hilbert space of q-bits at each site is $2^m$-dimensional, this
involves acting on the state space with a $2^m\times 2^m$ unitarity matrix.
We will restrict attention to collision operators which conserve
particle number.  

We have thus given a general formulation of the state space and
dynamics of a quantum lattice-gas automaton.  There is a very natural
parallel between the evolution of a single state of a QLGA and the
evolution of an {\it ensemble} of states in a stochastic
(nondeterministic) classical LGA.  In a stochastic LGA, the collision
rule acting on the states at a given lattice site is only defined
probabilistically.  We can define an ensemble of states by associating
a probability $p_{\sigma}$ with every state $\sigma$.  In a stochastic
LGA, the advection part of the time evolution rule permutes the
probabilities $p_\sigma$ by permuting the individual bits.  The
collision rule has the effect of acting on the vector of probabilities
$p_\sigma$ by a matrix which is probability-conserving, in the sense
that the columns as well as the rows of the matrix sum to unity.  The
only difference between this dynamics on an ensemble and the
definition given above of a QLGA is that the collision matrix for a
QLGA is unitary rather than probability-conserving.  Thus, we see that
at the expense of having exponentially more information contained in
each state, the QLGA naturally contains the complete dynamics of an
ensemble rather than those of a single instance.  As will be discussed
later, a QLGA can naturally be implemented on a quantum computer with
an exponential increase in performance.  In this case, the
measurements which must be performed to calculate results on the
quantum computer  correspond to single instances of
measurements in the physical system being simulated.  This highlights
one of the essential distinctions between a classical probabilistic
system and a quantum system, which is that the quantum
system contains information about all possible trajectories until such
a time as the system is measured.

\section{A simple example:  the free particle in 1D}
\noindent
Let us now begin by considering the simplest nontrivial QLGA we can
think of.  Consider a 1-dimensional lattice of size $l$ where each
lattice site has two possible occupation sites for particles ($m =
2$), corresponding to left- and right- moving particles.  At each time
step, the q-bits representing the particles at each lattice site will
hop one step to the left or right, and then the q-bits associated with
the new pair of particles at each lattice site will interact through a
collision matrix $T$.  If we restrict $T$ to conserve particle number
and we insist that $T$ be invariant under reflection, then in the
basis $--,+-,-+,++$ the matrix $T$ will be given (up to an irrelevant
overall phase) by
\begin{equation}
T = \left(\begin{array}{cccc}
1& 0 & 0 & 0\\
0 & q & p & 0\\
0 & p & q & 0\\
0 &0 &0 &  \phi
\end{array}\right)
\label{eq:k}
\end{equation}
where the parameters $q, p$ and $\phi$ satisfy $| q |^2 + | p |^2 =
1$, $p \bar{q} + \bar{p} q = 0$ and $| \phi |^2 = 1$.  Graphically,
the parameters $q$ and $p$ correspond to the amplitudes that a single
particle entering a lattice site will continue forward or bounce back
(see Fig.~\ref{f:bounce}).  The phase factor $\phi$ affects multiple
particle collisions, and will be discussed further later.
\begin{figure}[htbp]
\vspace*{13pt}
\centerline{\vbox{\hrule width 5cm height0.001pt}}
\begin{center}

\begin{picture}(280,110)(- 140,- 55)
\multiput(-60,0)(120,0){2}{\circle*{5}}
\put(-60,0){\vector( 1, -1){ 20}}
\put(60,0){\vector( -1, -1){ 20}}
\put(-80,20){\vector( 1, -1){ 17.5}}
\put(40,20){\vector( 1, -1){ 17.5}}

\put(-63,-30){\makebox(0,0){$q$}}
\put( 59,-30){\makebox(0,0){$p$}}
\put(-130, 10){\makebox(0,0){$t$}}
\put(-128,3){\vector( 0, -1){20}}

\end{picture}

\end{center}
\centerline{\vbox{\hrule width 5cm height0.001pt}}
\vspace*{13pt}
\fcaption{Collision rule for 1D QLGA}
\label{f:bounce}
\end{figure}

In order to understand the dynamics of this simple QLGA, let us
first restrict attention to the subspace of the total Hilbert space of
the system in which there is only a single occupied state.  Since the
collision rule conserves particle number, this subspace remains
invariant under the action of the time-development rule.  This
subspace is a $2l$-dimensional complex vector space, parameterized by
complex numbers $\psi_r (x), \psi_l (x)$ where $1 \leq x \leq l$.  The
QLGA dynamics defined above gives the equations of motion
\begin{eqnarray}
\psi_r (x, t + 1)& = &  q \psi_r (x -1, t)+ p \psi_l (x + 1, t) \nonumber\\
\psi_l (x, t + 1)& = &  q \psi_l (x +1, t)+ p \psi_r (x - 1, t) \label{eq:eom}
\end{eqnarray}

It is worth noting that the 1-particle system described here is very
similar to a discrete formulation of the 1+1D Dirac equation discussed
long ago by Feynman\cite{Fh}.  In Feynman's model a similar dynamics
is considered, however the parameter $p$ is taken to scale as
$\epsilon$.  The Dirac system was taken as the starting point in the
work of Meyer\cite{Meyer}.  The emergent \Schrodinger behavior of the
system with fixed $p$ was also discussed by Succi and Benzi\cite{sb}.

Just as the continuum behavior of a classical LGA can be determined by
performing a power series expansion of the equations of motion and
applying the Chapman-Enskog procedure\cite{lga,bw1}, we can take the
continuum limit of the equations of motion (\ref{eq:eom}) and
determine a set of differential equations satisfied by $\psi_r$ and
$\psi_l$ in the continuum limit.  Scaling
$x$ as $\epsilon$ and $t$ as $\epsilon^2$, we take the limit $\epsilon
\rightarrow 0$.  Factoring out a time-dependent phase factor from the
total amplitude
\begin{equation}
\Psi (x, t)= (p + q)^{-t}   (\psi_l (x, t) + \psi_r (x, t))
\label{eq:factor}
\end{equation}
we find\cite{bw2} that the total amplitude satisfies the \Schrodinger
equation
\begin{equation}
\frac{\partial}{\partial t} \Psi (x, t)
= \frac{i}{2m} \frac{\partial^2}{\partial x^2} 
\Psi (x, t)
\label{eq:Schrodinger}
\end{equation}
for a free particle of mass $m = ip/q$.  Note that the mass is real
because of the restriction $p \bar{q} + \bar{p} q = 0$.  The equation
(\ref{eq:Schrodinger}) can also be derived by mode analysis\cite{bw3}.

\section{The free particle in $D$ dimensions}
\noindent
We have seen that in the single-particle sector of a general 1D QLGA
the continuum limit gives a free \Schrodinger particle moving on a
line.  Let us now consider a QLGA in an arbitrary number of
dimensions, still restricting to the single-particle sector.  Let us
assume that we have a Cartesian lattice in $D$ dimensions, with $m
=2D$ q-bits at each lattice site, corresponding to particles moving
along any of the lattice vectors.  In the single-particle sector, the
collision rule is defined by a unitary $2D\times 2D$ matrix.  If we
assume that the collision rule is invariant under the symmetry group
of the lattice, we find that the set of allowed collision rules is
parameterized by 3 complex phases $\mu, \nu, \lambda$.  These phases
correspond to the eigenvalues of vectors in the 3 irreducible
representations of the discrete rotation group.  In particular, $\mu$
is the eigenvalue associated with the constant vector $(1, 1, \ldots,
1)$ and $\nu$ is the eigenvalue of vectors which change sign under a
parity transformation.  Just as in the 1D case, a systematic
analysis\cite{bw2} shows that as long as $\mu \neq \nu$ and $\mu \neq
\lambda$, the total amplitude after removing a phase
\begin{equation}
\Psi (x, t)= \mu^{-t}   \sum_{i}  \psi_i (x, t) 
\label{eq:factor2}
\end{equation}
satisfies the \Schrodinger equation
\begin{equation}
\frac{\partial}{\partial t} \Psi (\bfx, t)
= \frac{i}{2m} \sum_{i} \frac{\partial^2}{\partial x_i^2} 
\Psi (\bfx, t)
\label{eq:Schrodinger2}
\end{equation}
where the mass $m$ is related to $\mu$ and $\nu$ through
\begin{equation}
\frac{i}{2m}  =  \frac{1}{d}  \left(\frac{\nu}{\mu-\nu}  +\frac{1}{2}\right).
\label{eq:mass}
\end{equation}

As a simple concrete example of this general result, we can define a
collision rule in the single-particle sector of a $D$-dimensional QLGA
to have $\nu = 1, \lambda = -1$ with $\mu$ an arbitrary complex
phase.  With these phases, the amplitude for a particle to completely
reverse direction in the collision phase is given by
\begin{equation}
\alpha =\frac{\mu + 1 -D^2 -D}{D^2 + D}
\label{eq:a}
\end{equation}
while the amplitude for a particle to ``bounce'' in any other
direction is given by
\begin{equation}
\beta =\frac{\mu + 1}{D^2 + D}
\label{eq:b}
\end{equation}
These collision rules are described graphically in the case $D = 2$ in
Figure~\ref{f:rule2}.
\begin{figure}[htbp]
\vspace*{13pt}
\centerline{\vbox{\hrule width 5cm height0.001pt}}
\begin{center}
\begin{picture}(200,140)(- 100,- 55)

\multiput(-50,50)(100,0){2}{\circle*{5}}
\multiput(-50, -30)(100,0){2}{\circle*{5}}
\multiput(-70,50)(100,0){2}{\vector( 1,0){17.5}}
\put(-70,-30){\vector( 1,0){17.5}}
\put(30,-27){\vector( 1,0){17.5}}
\put(-50,50){\vector( 0,1){20}}
\put(50,50){\vector( 0,-1){20}}
\put(-50,-30){\vector( 1,0){20}}
\put(50,-33){\vector( -1, 0){20}}

\put(-52,20){\makebox(0,0){$\beta$}}
\put(48,20){\makebox(0,0){$\beta$}}
\put(-52,-43){\makebox(0,0){$\beta$}}
\put(48,-43){\makebox(0,0){$\alpha$}}

\end{picture}
\end{center}
\centerline{\vbox{\hrule width 5cm height0.001pt}}
\vspace*{13pt}
\fcaption{A QLGA collision rule in the single-particle sector giving a
\Schrodinger equation in 2D}
\label{f:rule2}
\end{figure}

With this choice of collision rule, in any dimension $D$ the total
amplitude $\Psi$ satisfies the Schrodinger equation
(\ref{eq:Schrodinger2}) with mass
\begin{equation}
m = i \frac{\mu -1}{d (\mu + 1)}.
\label{eq:mass2}
\end{equation}

\section{Adding an external potential}
\noindent

Thus far, the models we have considered describe in the
single-particle sector the propagation of a free \Schrodinger
particle.  We now consider the addition of an external potential.  In
order to incorporate the effects of a potential function $V (\bfx)$ we
need only multiply the wave function at each time step by an
overall phase factor of $\exp (-i \epsilon^2 V (\bfx))$ when there is
a particle at position $\bfx$.  In the QLGA framework, this
corresponds to transforming each q-bit by an operator which acts on the
Hilbert space (in the basis $-,+$) by the matrix
\begin{equation}
U =  \left(\begin{array}{cc}
1 & 0\\
0 &e^{-i\epsilon^2 V (\bfx)}
\end{array} \right)
\label{eq:potential}
\end{equation}
where $\bfx$ is the position associated with the given q-bit.  If we
modify the time-development rule for the QLGA so that after each
collision step each q-bit is acted on with this operator, the
resulting \Schrodinger equation in the single-particle sector is
\begin{equation}
\frac{\partial}{\partial t} \Psi (\bfx, t)
= \frac{i}{2m} \sum_{i} \frac{\partial^2}{\partial x_i^2} 
\Psi (\bfx, t)-iV (\bfx, t) \Psi (\bfx, t).
\label{eq:Schrodinger3}
\end{equation}

\begin{figure}[htbp]
\vspace*{13pt}
\centerline{\vbox{\hrule width 5cm height0.001pt}}

\vskip -2.85cm
\centerline{
\psfig{figure=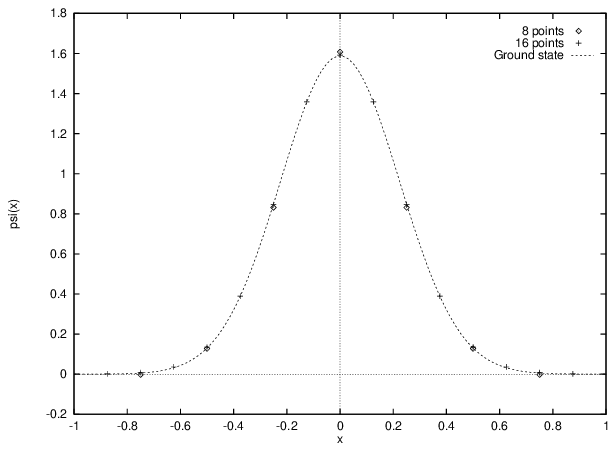,angle=0,height=13.5cm}}
\vskip -3.4cm

\centerline{\vbox{\hrule width 5cm height0.001pt}}
\vspace*{13pt}
\fcaption{Ground state in quadratic
potential with 8 and 16 lattice sites}
\label{f:function1}
\end{figure}

\begin{figure}[htbp]
\vspace*{13pt}
\centerline{\vbox{\hrule width 5cm height0.001pt}}

\vskip -2.85cm
\centerline{
\psfig{figure=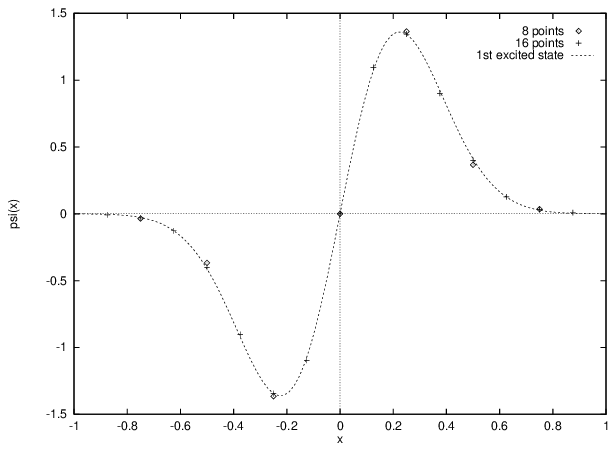,angle=0,height=13.5cm}}
\vskip -3.4cm

\centerline{\vbox{\hrule width 5cm height0.001pt}}
\vspace*{13pt}
\fcaption{First excited state in quadratic
potential with 8 and 16 lattice sites}
\label{f:function2}
\end{figure}

As an example of this type of system, let us consider a single
particle moving in one dimension in a harmonic oscillator potential $V
(x)= x^2/2$.  We can combine the advection operator, the collision
matrix $T$ from (\ref{eq:k}) at each point, and the external potential
(\ref{eq:potential}) at each point into a single time-development
matrix which acts on the Hilbert space.  Diagonalizing this matrix
gives the eigenstates of the time-development equation, which should
approximate the energy eigenfunctions of the corresponding quantum
system in the limit as the lattice spacing becomes small.  As a test
of the method, we have analyzed this system numerically on small
lattices (a similar analysis for a square well potential was performed
by Meyer\cite{Meyer2}).  Even for very small lattices, we find that
the first few eigenfunctions are extremely close to the wavefunctions
of the continuous theory.  In Figures~\ref{f:function1}
and~\ref{f:function2} we have graphed the ground state and first
excited states for lattices with 8 and 16 lattice sites (we have only
included effects of every second site since particles at sites of
opposite parity never interact).  As can be readily seen in the
graphs, even with only 8 lattice sites the first two eigenstates are
reproduced very accurately by this discretization.  As the size of the
lattice increases, the number of eigenstates of the continuum system
which are correctly reproduced increases proportionally.

\section{Many-body systems}
\noindent
Now that we understand the behavior of the single-particle sector of a
QLGA, let us return to the more general situation where there are $n$
particles in the system.  As long as $n\ll lm$, the occupied particle
sites will be sparse in the system.  At
most lattice sites where particles are present there will only be a
single particle, so that each particle will independently satisfy a
\Schrodinger equation.  Only when several particles arrive at a
lattice site at the same time will the particles interact.  In this
case, the part of the collision rule in the multiple particle sector
will define a local (delta function) interaction between the
particles.  Thus, in the $n$-particle sector of the QLGA, the
continuum limit of the dynamics will be a system of $n$ particles
moving according to the $n$-body \Schrodinger equation with delta
function interactions.

As an example of this type of system, we can describe the collision
matrix for a gas of nonrelativistic particles interacting by local
($\delta (x-y)$) interactions in an external quadratic potential $V
(x) = ax^2$ in one dimension.  Incorporating the potential term into the
collision matrix, we have
\begin{equation}
T = \left(\begin{array}{cccc}
1& 0 & 0 & 0\\
0 & \frac{\mu +1}{2}\left( e^{-i a\epsilon^2 x^2}\right) &
\frac{\mu -1}{2}\left( e^{-i a\epsilon^2 x^2}\right) & 0\\[0.1in]
0 & \frac{\mu -1}{2} \left(e^{-i a\epsilon^2 x^2}\right) &
 \frac{\mu +1}{2}\left(e^{-i a\epsilon^2 x^2}\right) & 0\\
0 &0 &0 & \phi e^{-2 i a\epsilon^2 x^2} 
\end{array}\right)
\label{eq:gas}
\end{equation}
where $\phi$ is a complex phase determining the effects of the delta
function interaction between particles.

Just as we can incorporate an external potential by rotating each
q-bit by an appropriate phase, we can incorporate an arbitrary
interparticle potential by acting on each pair of q-bits.  Given an
arbitrary function $V (\bfx,\bfy)$ describing an interaction potential
between particles at positions $\bfx,\bfy$, we can act on each pair of
q-bits at each time step with the matrix
\begin{equation}
U = \left(\begin{array}{cccc}
1& 0 & 0 & 0\\
0 & 1 & 0 & 0\\
0 & 0 & 1 & 0\\
0 &0 &0 & e^{- i \epsilon^2 V (\bfx, \bfy)} 
\end{array}\right).
\label{eq:interaction}
\end{equation}
By including such an interaction in the time development of the QLGA,
we can model {\it any} interacting nonrelativistic quantum many-body
system of interest.

It is worth noting that the systems we have described are most easily
used to simulate a system of nonrelativistic bosons, despite the
apparent use of exclusionary statistics.  The simulation of fermions
is also possible, however some extra overhead is necessary for keeping
track of relative phases\cite{al}.

\section{Computational complexity}
\noindent
We have described a class of algorithms which can be used to simulate
an arbitrary system of interacting nonrelativistic quantum particles.
We will now discuss briefly the computational complexity of these
algorithms, for implementations on both classical and quantum
computers.

Let us consider a system of $n$ particles moving on a $D$-dimensional
lattice of size $l = q^D$, with $m = 2D$ allowed particle positions
per lattice site.  Assuming that $n\ll lm$, the number of complex
variables needed to describe the
state of the system at a point in time is
\begin{equation}
\frac{(lm) !}{n !(lm-n) !}  \sim \frac{(lm)^n}{n!}.
\label{eq:variables}
\end{equation}
To simulate a system with this number of variables on a classical
computer would take at the very least on the order of the number of
variables for each time step.  The number of time steps needed scales
as $q^2$ (because $t$ scales as $\epsilon^2$), so the total time
needed for a computation on a classical computer would be
\begin{equation}
T_c \approx{\cal O} (\frac{ q^{2 + Dn}m^n}{n!} )
\label{eq:tc}
\end{equation}
For a typical calculation of physical interest, we might have $n =
100, q = 20, D = 3$.  For such a calculation, the number of operations
needed on a classical computer would be on the order of $T_c \approx
10^{312}$.  This is clearly impractical.  Note that a standard finite
difference method would eliminate the factor of $m^n$ in
(\ref{eq:tc}), however this would not make such a calculation any more
accessible.  Only when the number of particles $n$ is extremely small
is it conceivable that these algorithms might be a useful approach for
simulating quantum systems on a classical computer.

Now let us consider the computational complexity of the same
algorithms on a quantum computer.  For the quantum simulation, we need
$m\cdot l$ q-bits.  The local advection and collision steps can be
accomplished with on the order of $m\cdot l$ quantum operations per time
step.  Thus, a system of quantum particles which affect one another
only through local delta function interactions can be simulated on a
quantum computer in time on the order of
\begin{equation}
T_q \approx{\cal O} (2Dq^{2 + D}).
\label{eq:tq}
\end{equation}
Note that this time is {\it independent} of the number of particles
$n$ being simulated.  In fact, this algorithm will simultaneously
simulate the system for all allowed numbers of particles $n \leq lm$
in the same time it takes to simulate a system with only a single
particle.  With the numbers used in the example above on a classical
machine, the number of operations needed to perform the simulation is
a much more tractable $T_q \approx 19.2\cdot 10^6$.  The idea that it might
be possible to simulate quantum mechanical systems exponentially
faster on a quantum computer than on a classical computer was first
suggested by Feynman\cite{Feynman}; a general argument for this
conclusion was given more recently by Lloyd\cite{Lloyd}.  The
algorithms discussed in this talk represent a concrete instantiation
of the general principles discussed by those authors.

An additional factor appears in the computational complexity of the
algorithm when we have an arbitrary interparticle potential.  Because
in this case at every time step we must include an operation for every
pair of q-bits, the complexity on a quantum computer becomes
\begin{equation}
T_q \approx{\cal O} (4D^2q^{2 + 2D}).
\label{eq:tq2}
\end{equation}
In the example discussed above, this increases the computational
complexity to $T_q \approx 10^{12}$ operations.  Clearly, a fairly
sizable quantum computer would be needed to carry out such a
calculation.  Note, however, that by comparison a standard home PC can
currently perform this number of operations in about 15 minutes.

\section{Conclusions}
\noindent
In this paper we have described a class of discrete algorithms for
simulating the many-body \Schrodinger equation.  Under fairly simple
conditions of isotropy and genericity, the  behavior of a general
quantum lattice-gas automaton which preserves particle number is to
simulate a many-body Schrodinger equation with pointlike interactions.
By adding a nonlocal interaction term, we can simulate any interacting
nonrelativistic quantum system of interest using quantum lattice-gas
automata models.

Because the number of degrees of freedom in the quantum system is so
large, it is impractical to use the algorithms described here on a
classical computer to simulate more than a handful of interacting
particles.  Because simulating quantum systems is such a difficult
problem, however, these algorithms may be useful in certain situations
even on a classical computer, due to their inherent unitarity.

The real utility of these algorithms will be realized only if they can
be implemented on quantum computers.  At the moment, it is rather
unclear whether a general purpose quantum computer capable of
performing millions or billions of coherent quantum operations can be
constructed, even in principle.  There are a number of serious
technical challenges to be overcome in constructing such a system.
There are also possible theoretical obstacles due to decoherence
problems.  Recent work has indicated that decoherence and imprecision
problems can be overcome by clever use of quantum error correction
codes\cite{correction}.  Nonetheless it will be many years before a
working quantum computer of reasonable size will be available, even if
all the technical problems can be solved.

If, however, there ever are general purpose quantum computers
available for use in scientific research, the algorithms described
here would allow for the simulation of a wide range of quantum systems
of physical interest which are inaccessible to simulation on classical
computers.  Using quantum lattice-gas automata, any interacting
nonrelativistic quantum many-body system could be simulated.  This
would allow for the study of systems including electron gases, metals,
plasmas, nuclear matter, Fermi gases, and many other phenomena of
physical and industrial interest.

In this paper we have only discussed simulations of nonrelativistic
\Schrodinger systems using QLGA.  There are other more complicated
physical systems which may also be accessible using these same
methods.  Some work has been done\cite{Meyer} on simulating a
many-body Dirac equation using QLGA.  A class of systems for which it
would be particularly interesting to find QLGA models are abelian and
nonabelian gauge theories.  In particular, there is a fairly large
research effort devoted to the numerical study of quantum
chromodynamics (QCD), the nonabelian gauge theory coupled to fermions
which describes the interaction of quarks\cite{Creutz}.  If it were
possible to simulate QCD using a simple QLGA lattice model, this would
indicate that quantum computers could be used to simulate QCD in the
Hamiltonian framework\cite{ks} with an exponential speedup, possibly making
accessible to numerical experiment a number of poorly understood
aspects of this important theory.

\nonumsection{Acknowledgements}
\noindent
We would like to acknowledge helpful conversations with Francis
Alexander, Peter Coveney, Eddie Farhi and Jeffrey Yepez.  BMB was
supported in part by Phillips Laboratories and by the United States Air
Force Office of Scientific Research under grant number F49620-95-1-0285.
WT was supported in part by the divisions of Applied Mathematics of the
U.S.  Department of Energy (DOE) under contracts DE-FG02-88ER25065 and
DE-FG02-88ER25066, in part by the U.S. Department of Energy (DOE) under
cooperative agreement DE-FC02-94ER40818, and in part by the National
Science Foundation (NSF) under contract PHY90-21984.

\nonumsection{References}


\noindent

\begin{thebibliography}{000}

\bibitem{qc} For recent reviews of quantum computation, see
A.\ Ekert and R.\ Jozsa, {\bibit Rev.\ Mod.\ Phys.} {\bf 68}, No.\ 3,
733 (1996); S.\ Lloyd, {\bibit Sci.\ Amer.}, 140 (Oct.\ 1995);
D.\ P.\ DiVincenzo, {\em Science} {\bf 270} 255 (1995).


\bibitem{Meyer}  D.\ Meyer, ``From quantum cellular automata to
quantum lattice gases,'' UCSD preprint,  
{\tt quant-ph/9604003}, March 1996.
\bibitem{lga} U.\ Frisch, B.\ Hasslacher and Y.\ Pomeau, {\bibit Phys. Rev.
   Lett.} {\bf 56}, 1505 (1986); U.\ Frisch,  D.\ d'Humi\`{e}res,  B.\
   Hasslacher, 
    P.\ Lallemand,  Y.\ Pomeau and J.-P.\ Rivet, {\bibit Complex Systems} {\bf
   1},  648-707  (1987);  S.\ Wolfram, {\bibit J. Stat. Phys.}, {\bf 45},
   471 (1986).
\bibitem{Fh}  R.\ P.\ Feynman and A.\ R.\ Hibbs, , {\bibit Quantum Mechanics
   and Path Integrals} (McGraw-Hill, New York, 1965) pp. 35-36; see
   also Feynman's unpublished notes as reproduced in S.\ S.\ Schweber,
   {\em Rev. Mod. Phys.} {\bf 58}  449 (1986).
\bibitem{sb} S.\ Succi and R.\ Benzi, {\bibit Physica} {\bf D69}, 327
(1993); S.\ Succi, ``Numerical solution of the Schr\"odinger equation
using discrete kinetic theory'', IBM-ECSEC preprint, 1995.
\bibitem{bw1} B.\ Boghosian and W.\ Taylor, ``Correlations and
Renormalization in Lattice Gases'', 
Phys.\ Rev.\ {\bf E} 52, 510-554 (1995).
\bibitem{bw2} B.\ Boghosian and W.\ Taylor, ``A quantum lattice-gas
model for the many-body Schr\"odinger equation in $d$ dimensions'',
BU-CCS/PUPT preprint {\tt quant-ph/9604035}, April 1996.
\bibitem{bw3} B.\ Boghosian and W.\ Taylor, ``Simulating quantum
mechanics on a quantum computer'', BU-CCS/PUPT preprint, January 1997.
\bibitem{Meyer2} D.\ Meyer, ``Quantum mechanics of lattice gas
automata I.  One particle plane waves and potentials'', UCSD preprint
{\tt quant-ph/9611005}, October 1996.
\bibitem{al} D.\ S.\ Abrams and S.\ Lloyd ``Simulation of Many-Body Fermi
Systems on a Universal Quantum Computer'', MIT preprint, November
1996.
\bibitem{Feynman} R.\ Feynman
{\bibit Int.\ J.\ Theor.\ Phys.} {\bf 21}, 467-488 (1982); {\bibit Found.\
Phys.} {\bf 16}, 507-531 (1986).
\bibitem{Lloyd}  S.\ Lloyd, {\it Science} {\bf
    273}, 1073 (1996).
\bibitem{correction} P.\ W.\ Shor, ``Fault-tolerant quantum
computation'', AT\&T preprint\\ {\tt quant-ph/9605011}, May 1996, and
references therein.
\bibitem{Creutz} See for example, M.\ Creutz, {\bibit Quarks, Gluons, and
Lattices} (Cambridge U.\ Press 1983). 
\bibitem{ks}  J.\ Kogut and L.\ Susskind, {\bibit Phys.\ Rev.} {\bf
D11}, 395 (1975).


\end{thebibliography}
\end{document}